# A web-tool for calculating the economic performance of precision agriculture technology

M. Medici[1], S.M. Pedersen[2], M. Canavari[1]*, T. Anken[3], P. Stamatelopoulos[4], Z. Tsiropoulos[4], A. Zotos[4], G. Tohidloo[3]

[1]Alma Mater Studiorum-Università di Bologna, Viale Giuseppe Fanin 50, 40127 Bologna, Italy

[2]University of Copenhagen, Rolighedsvej 25, 1958 Frederiksberg, Denmark

[3]Agroscope, Tänikon, 8356 Ettenhausen, Switzerland

[4]Agenso, Markou Mpotsari 47, 117 42 Athens, Greece

*corresponding: maurizio.canavari@unibo.it, Department of Agricultural and Food Sciences, Alma Mater Studiorum - Università di Bologna. Address: viale Giuseppe Fanin 50, 40127 Bologna. Tel. +39 051 2096108

# A web-tool for calculating the economic performance of precision agriculture technology

Abstract

To develop precision agriculture (PA) to its full potential and make agriculture progress toward sustainability and resilience, appropriate criteria for the economic assessment are recognised as being one of the most significant issues requiring urgent and ongoing attention. In this work, we develop a web-tool supporting the assessment of the net economic benefits of integrating precision farming technologies in different contexts. The methodological approach of the tool is accessible to any agricultural stakeholder through a guided process that allows to evaluate and compare precision agriculture technologies with conventional systems, leading the final user to assess the financial viability and environmental impact resulting from the potential implementation of various precision agriculture technologies in his farm. The web-tool is designed to provide guidelines for farmers over their decisions to invest in selected PA technologies, by increasing the knowledge level about novel technologies characteristics and the related benefits. Possible input reduction also offers the possibility to investigate the mitigation of environmental impacts.

**Keywords**: *precision agriculture (PA), technology, adoption, cost-benefit analysis, economic performance, financial analysis, sustainability, web application*

## 1. Introduction

This study describes the development of an on-line web-tool, intended as a prototype application, for calculating the economic performance of precision agriculture (PA) technology within the European research project PAMCoBA (Precision Agriculture Methodologies for Cost-Benefit Analysis). Farmers as well as any agricultural stakeholder can freely access the proposed methodology through a guided process that allows to evaluate and compare profitability associated to PA technologies, assessing the financial viability derived from the potential adoption of various PA systems.

Precision agriculture (PA) is a farm management system involving crop management based on field variability and site-specific conditions (Seelan et al., 2003). It includes a number of technologies ranging from sensing systems that map crop and soil variability to guidance systems and variable-rate (VR) systems that dose agricultural inputs onto the field. These technologies have a wide potential to improve agricultural performance, ranging from more efficient crop nutrient use, increased crop quality and quantity, and reduced field overlaps. For any of these reasons, PA is generally associated to potential economic and environmental benefits (Li et al., 2018).

Although the implementation of new technologies is essential for farmers to remain competitive in their business, PA is a complex management system requiring changing from empirical decision-making towards data-driven decision processes, with benefits hard to quantify in advance. As a result, the adoption of PA technology among farmers remain low in Europe (STOA, 2016). In recent years, many authors have stressed the uncertainty regarding the non-clear perception of benefits when using PA technologies, as well as the potential cost savings and reduction in environmental impacts resulting from their adoption (Eastwood et al., 2016; Kutter et al., 2011; McBratney et al., 2005; Medici et al., 2019).

It appears that the large knowledge gap between farmers and technology developers results in difficulties in explaining the economic and environmental benefits characterizing PA. In this context, the successful adoption of PA technologies requires efforts and confidence from farmers, who need to continuously integrate new knowledge (Oreszczyn et al., 2010), and this is likely to contribute in mitigating uncertainty about potential benefits arising from PA. The development of integrated performance assessments, which could influence the rate of dissemination among farmers too, may constitute a possible solution to the issues above described (Lamb et al., 2008). For these reasons, appropriate economic criteria for technology adoption requires urgent and ongoing attention in order to develop PA to its full potential.

The web-tool was designed within the context of the ICT-Agri project '*PAMCoBA*' (*Precision Agriculture Methodologies for Cost Benefit Analysis*) to provide guidelines for farmers over their decisions to invest in selected precision agricultural technologies. It is freely available at: http://tool.pamcoba.eu/. The tool explores data regarding existing PA technologies, crops and agricultural operations, guiding farmers in the selection of the most appropriate technologies for farm specific context. It is particularly suitable for arable crops, with wheat, , sugar beet, canola, and potato, modelled, but it also offers the possibility of modelling any other custom crop, by inserting values defining yield, price, and agricultural treatments. As a final result, the tool can evaluate the profitability, supporting farm decision-making processes.

## 2. Web-tool structure and database

The web-tool was designed according to the latest trends on software development for the World Wide Web. The scripting language used was PHP (Hypertext Pre-processor), and specifically, the Laravel framework, which is the most popular tool for developing fast and stable web applications, was adopted. The database is hosted on a MySQL server instance and it was designed according to best practices to maintain data integrity and security and empowered with custom scripts and functions to enable automated backup mechanisms.

Users interact with the webtool and perform their analysis through the user interface that was developed using the VueJS framework, which is one of the most popular JavaScript frameworks enabling the creation of fast and reliable applications.

The design of the webtool resides in the interplay of global data structured in a relational database at administration (server) level in which connections between crops, agricultural activities (following named as 'operations') and technologies are managed. Moreover, in this context information regarding costs related to operations and underlying crops (e.g. seed, fertilizer and pesticide prices, no. treatments per year, fuel price, fuel consumption per hectare, labour cost and labour hours required per hectare) and PA technology costs are defined. The final user can adjust this auto-fill data and in any case, he is required to define entry values regarding farm-level information (e.g., location, farm size, yields, prices and selected crops), as explained in Section 3.

In the web-tool the following operations have been outlined: seeding, fertilization (i.e. application of mineral nitrogen), spraying (i.e. application of pesticides – fungicides, herbicides, insecticides, and growth regulators), mechanical weeding, tillage, liming and manure application.

Consequently, a broad number of PA technologies able to perform these activities were integrated: auto-steer, section control, VR fertilization system, VR seeding system, VR spraying system, VR liming system, VR manure application system, inter-row hoeing system (with camera or GPS).

Also, several support technologies were additionally considered to be coupled with the aforementioned technologies, in order to give the chance to increase the expected performance of the PA systems in the field; the support technology includes geolocation systems (normal GPS antenna, RTK-GPS antenna, or satellite), controlled traffic farming (CTF), surveying Unmanned Aerial Vehicles (UAVs), N-sensor, yield mapping systems, and devices for the measurement of soil electrical conductivity (EC) and soil pH.

The webtool allows users to retrieve information about each possible technology combination useful to assess the related economic performance.

## 2.1 Benefits of PA technologies

Modelled benefits for PA technologies are agricultural input savings, yield increase, fuel saving and labour saving. Most reference values have been quantified according to evidences from scientific and technical literature. Missing values have been assessed by the authors based on their experience and common sense, staying on the safe side with conservative estimates. In this work, we mostly refer to literature reviews and analysis recently performed by the Authors (Medici et al., 2019; Pedersen et al., 2019). As numerous technologies are included in the tool and this study focus on the methodological approach of the web-tool, only the most relevant benefits regarding PA technologies are discussed in the following. Table 1 reports the suggested "default" benefits for each technology system modelled in the web-tool, which in most cases are based on conservative estimations compared to those available in the literature. Nonetheless, the original idea at the basis of the tool is that the user can take the proposed library as a reference and adjust/insert new values based on his knowledge and experience.

*Auto steer & CTF*. The use of auto-steer can lead to overlap reductions. For instance, Batte & Ehsani (2006) found a 3-7% overlap reduction for auto-guidance systems, and then observed that the saving of agricultural inputs increase proportionally to un-overlapped area. Another work indicates that the reduced overlap with auto-steering can reach about 5-10% (Petersen et al., 2006). Conversely, one study has shown no relevant effect in terms of input saving (Holpp et al., 2013).

*Section control*. Automatic section control on boom sprayers can reduce overlaps when turning on headlands, and this feature is particularly significant when fields are irregularly shaped. There is a potential for reduced overlap of up to 5% for section control when using a sprayer but it might be smaller in case of dry granular fertilizer products The real potential is case bounded, as field shape and headland size can vary a lot (Pedersen and Pedersen 2018).

*VR applications*. Regarding VR fertilization, it is documented that lower rates of nitrogen fertiliser carefully applied can reduce fertiliser use without affecting yield and crop quality in maize (Schmidt et al., 2002). In cereal crops Bourgain & Llorens (2009) reported a 11.1% saving of nitrogen, while (Casa et al., 2011) reported a 22% reduction in N used in comparison to uniform applications, with no relevant effect on yields. A number of approaches have been developed to produce prescription maps for site-specific applications generated via satellite, survey UAV or through a Yara N-sensor mounted on tractor.

The adoption of VR spraying is associated to agronomic benefits too. Dammer & Adamek, (2012) estimated a 13.4% of insecticide saving in cereals; regarding fungicide application, (Pedersen and Pedersen 2018) documented a 1% increase in crop yield.

Economic benefits from VR lime application strongly depend on crop and soil (e.g. soil acidity and homogeneity) and optimal lime rates (Pedersen, 2003). A couple of studies reported yield benefits regarding VR lime application: (Weisz et al., 2003) explored the long-term impact of site-specific lime application on a soybean field reporting significant increases in soil pH and higher yields (up to 14%), while (Bongiovanni & Lowenberg-Deboer, 2000) observed yields increasing between 1.8% and 4.8% in maize and soybean.

To date, there is poor evidence regarding the economic impact of VR manure application. Lack of accuracy, basically consisting in nutrient overlaps, and nutrient content information are present issues; the nutrient content can be monitored via near-infrared spectroscopy (NIR) but this technology is still in its infancy. In the web-tool, reference values for input reduction within 4% were considered, depending on the type of technology system used.

***Weeding.*** Since weed control can be performed mechanically or with herbicide application, it would be hard to model any input reduction; in this work we refer to a 50% labour saving as standard benefit. This parameter is intended as compared to a hoe steered by an additional operator.

In the web-tool, the most relevant precision agriculture technologies for European farmers are considered, including monitoring technologies, guidance technologies and VR technologies. Often monitoring technology providers offer services in addition to physical equipment, like software applications able to convert sensing data into readable parameters. For this reason, we have intended additional services like zone sampling and prescription maps as included in the monitoring technology already defined in the tool. In any case, the user can model more accurately this issue, choosing an option and then inserting the related costs and benefits.

*Table 1 - Default benefits for PA technologies*

| Option name | Main technology | Support technology | Input reduction[1] (%) | Yield increase (%) | Fuel reduction (%) | Labour reduction (%) |
|---|---|---|---|---|---|---|
| **1 Auto steer & CTF** | Auto-steer | Normal GPS | 3 | - | 3 | 1 |
| | | RTK-GPS | 3 | - | 3 | 1 |
| | | RTK-GPS, CTF | 3 | 1 | 5 | 1 |
| **2 Section control** | Section control sprayer/seeder/fertilizer | Normal GPS | 2 | - | - | - |
| | | RTK-GPS | 4 | - | - | - |
| **3 VR applications** | VR Seeder | Satellite | 3 | - | - | - |
| | | Survey UAV | 3 | - | - | - |
| | | Yield map | 3 | - | - | - |
| | | Soil EC | 3 | - | - | - |
| | VR Fertilizer | Satellite | - | 3 | - | - |
| | | Survey UAV | - | 3 | - | - |
| | | Yield map | - | 3 | - | - |
| | | Soil EC | - | 3 | - | - |
| | | N sensor | 1 | - | - | - |
| | | N sensor, yield map | 1 | 3 | - | - |
| | | N sensor, yield map, soil EC | 3 | 3 | - | - |
| | VR Sprayer (fungicide) | Satellite | 15 | - | - | - |
| | | N-sensor | 15 | - | - | - |
| | | Survey UAV | 20 | - | - | - |
| | VR Sprayer (insecticide) | Satellite, yield map, soil EC | 15 | - | - | - |
| | VR Sprayer (herbicide) | Satellite | 15 | - | - | - |

|  |  | Survey UAV | 15 | - | - | - |
|  |  | Survey UAV, Yield map | 20 | - | - | - |
|  | VR Sprayer (growth regulator) | Satellite | 15 | - | - | - |
|  |  | Survey UAV | 20 | - | - | - |
|  | VR Lime | Satellite, yield map, soil EC | 2 | 1 | - | - |
|  |  | Survey UAV, yield map, soil EC | 2 | 1 | - | - |
|  |  | N-sensor, yield map, soil EC | 2 | 1 | - | - |
|  | VR Manure | Satellite | 1 | - | - | - |
|  |  | Satellite, yield map | 2 | - | - | - |
|  |  | Satellite, yield map, Soil sampling | 3 | - | - | - |
|  |  | Survey UAV | 2 | - | - | - |
|  |  | Survey UAV, yield map | 3 | - | - | - |
|  |  | Survey UAV, yield map, Soil sampling | 4 | - | - | - |
| **4 Weeding** | Weed control | Inter-row hoeing with GPS | - | - | - | 50 |
|  |  | Inter-row hoeing with camera | - | - | - | 50 |

[1]Note: Input reduction refer to the input mentioned in column 2 for options 3 and 4. For options 1 and 2 input reductions refer to all inputs of seed, fertilisers, and pesticides.

## 3. User interface and application features

The user is guided in the PA technology assessment through an easy three-step procedure through: input data specification (*MY FARM*), definition of the list of operations and suitable PA technologies (*OPTIONS/TECHNOLOGY*) and potential benefit evaluation with final calculation (*ECONOMIC BENEFITS*).

### 3.1 Input data, operations and technologies

In the first step in the web-tool homepage the language and the login, as well as default values of different countries can be chosen, as outlined from Figure 1. Default values are differentiated
by geographical area (Northern Europe, Central Europe, South & Southwestern Europe, Southeast Europe) and by crop (Wheat, Maize, Sugar beet, Canola, Potato). The web app user can fill data regarding current yield (t/ha) for each selected crop, total crop surface (ha) and price (€/t).

User can select one or more options from the second screen (Figure 2), then suitable PA technologies can be selected according to the chosen operations, including controlled-traffic farming (CTF), if applicable.

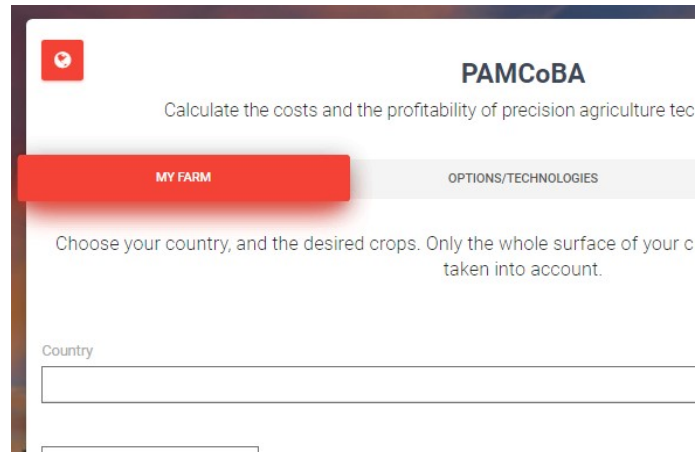
Figure 1 - PAMCoBA web-tool, MY FARM interface

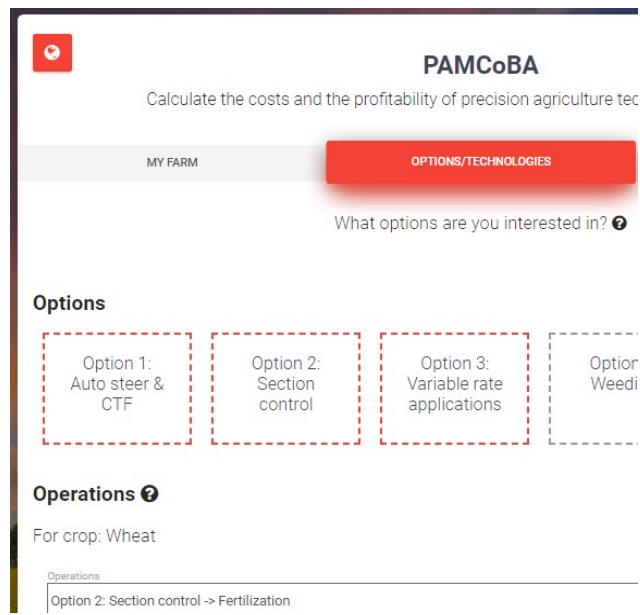
Figure 2 - PAMCoBA web-tool, OPTIONS/TECHNOLOGY interface

## 3.2 Evaluation of economic benefits

In the next screen, the proposed costs and benefits for each crop, options and technology system can be checked and changed, if needed (Figure 3). The user can also adjust the applied discount rate (interest rate) associated to the modelled investment in PA technologies, as well as the investment costs associated to each suitable PA technology made available from chosen crop and operations.

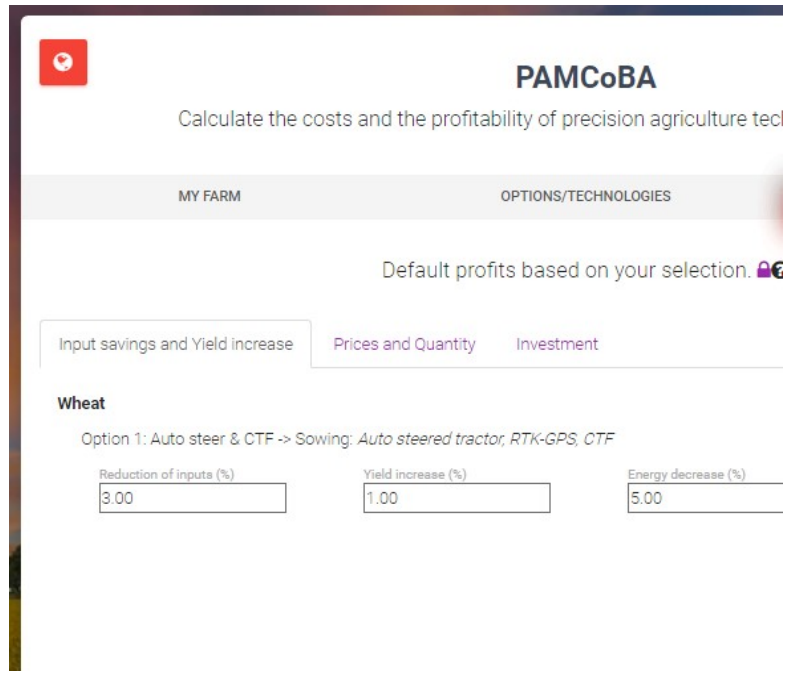

*Figure 3 - PAMCoBA web-tool, ECONOMIC BENEFITS interface*

The financial analysis embedded in the web-tool is based on the estimation of the differential cash flows from selected PA technologies with a description of initial investment costs *I* (€), input costs *C* (€/year) and expected benefits in the form of potential revenues *R* (€/year).

The proposed investment costs are thought to be appropriate for a farm up to 50 hectares. Thus, when the crop surface is increased, investment cost should be adequately re-adapted. It was decided to apply the "0.6 rule" to consider economies of scale (Tribe & Alpine, 1986), which origins from the relationship between the increase in equipment cost (i.e. investment) and the increase in capacity (i.e. crop surface). For more details, please refer to (Pedersen et al., 2019).

Benefit calculation related to each chosen operation was modelled adopting the Net Present Value approach (NPV) with cash flows that are modelled throughout a set period (in this case 8 years) as described by Equation 1:

$$NPV = -I' + \sum_{t=1}^{8} \frac{R_t - C_t}{(1+r)^t}$$

*Equation 1*

In which *I'* (€) is the total investment cost associated to the chosen operation scaled with the "0.6 rule", $R_t$ (€/year) are operational revenues equal to potential yield increase multiplied for current yield and crop price, $C_t$ (€/year) is the sum of costs at time *t* such as input, fuel and labour reduction, *r* is the discount rate. The internal rate of return (IRR) and the benefit-to-cost ratio (BCR) are also calculated for each selected system. By doing so, the tool allows to integrate several technologies at the same time and thereby integrate the synergies from applying, for instance, RTK-GPS systems for different operations and fields, as would be the case on real farms.

Figure 4 shows the web-tool flow chart for benefit calculation. Finally, the web-tool allows user to download a pdf report including the detailed economic performance of PA technology system suitable for the farm, as well as the information about the quantity of input saved.

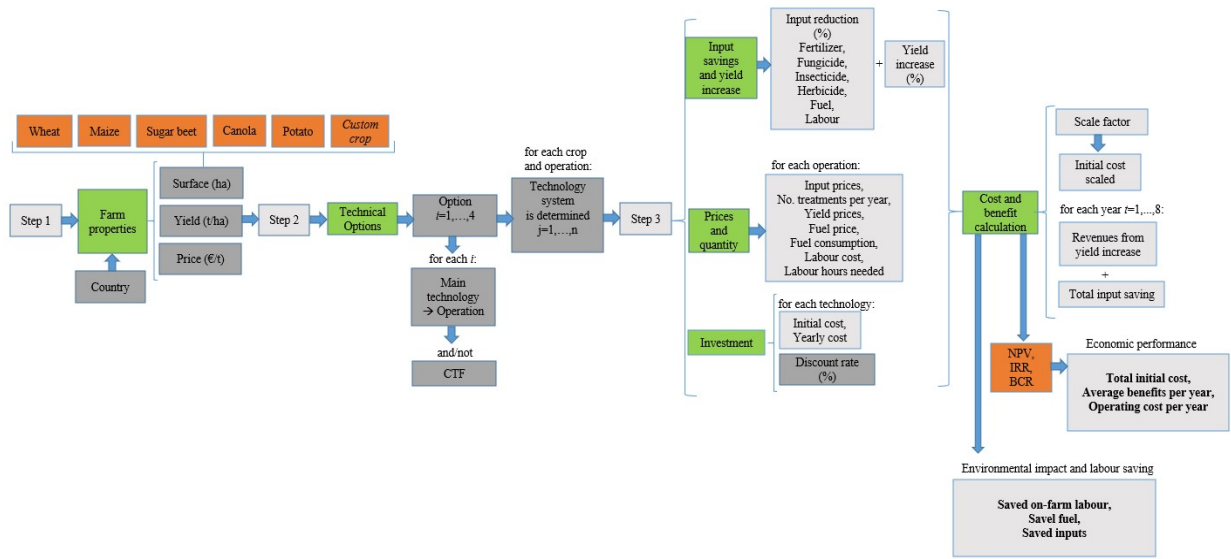

*Figure 4 – Flowchart for benefit calculation of PAMCoBA web-tool*

## 3.3 Additional features

The web-tool allows the user to choose initial parameters recommended for those unfamiliar with the production impacts of precision technologies. It foresees to adjust all input data needed to calculate economic benefit as well as the possibility to build-up a custom crop with their own values to be specified. As the potential benefits and savings for site-specific application may vary depending on specific locations, users can freely save and compare multiple runs, adjusting operations and parameters in order to have a complete overview of potential PA technology performance.

## 4. Conclusions

The development of PAMCoBA web-tool focused in managing data for the benefit of farmers. It was designed to provide guidelines for farmers over their decisions to invest in selected PA technologies, by increasing the knowledge level about novel technologies characteristics and the related benefits. We believe that considering and evaluating investments in agriculture is a good management practice, and if any farmer would manage investments in the best way possible, then considerations regarding the profitability of technology should be considered.

To our knowledge, this tool is unique, capable of helping stakeholders to understand the economics of PA technologies in an easy and friendly way. Its increased modifiability and adaptability, as well as the capability of allowing users to enter their specific data and to re-run the analysis by adjusting the system's values, can help users to understand how each technology can affect their farms positively. Moreover, the tool is novel as it integrates the application of multiple PA technologies in a whole farm management perspective. A powerful database was developed to achieve a high level of user customization, allowing the modification of the tool and its parameters through its administration panel.

The information about the quantity of input saved, reported in the downloadable report, may also be used outside this context, for instance to model environmental assessments scenarios. The tool can also be potentially useful to policy-makers if many farmers use it and provide (anonymous) data suitable to do policy analysis.

Appropriate criteria for the economic assessment of novel technology adoption is recognized as being one of the most significant issues requiring urgent and ongoing attention in order to develop PA to its full potential. We hope that this web-tool can help to improve the adoption rate of PA technologies in the future.

## Acknowledgement

This project has received funding from the European Union's Seventh Framework Programme for research, technological development and demonstration under grant agreement no 618123 ERA-NET - ICT-Agri in the project PAMCOBA.